# Negative Space: Workspace Awareness in 3D Face-to-Face Remote Collaboration

**Maurício Sousa, Daniel Mendes, Rafael K. dos Anjos, Daniel Simões Lopes, and Joaquim Jorge**
INESC-ID Lisboa, Instituto Superior Técnico, Universidade de Lisboa
{antonio.sousa, danielmendes, rafaelkuffner, daniel.s.lopes, jorgej}@tecnico.ulisboa.pt

## ABSTRACT
Face-to-face telepresence promotes the sense of "being there" and can improve collaboration by allowing immediate understanding of remote people's nonverbal cues. Several approaches successfully explored interactions with 2D content using a see-through whiteboard metaphor. However, with 3D content there is a decrease in awareness due to ambiguities originated by participants' opposing points-of-view. In this paper we investigate how people and content should be presented for discussing 3D renderings within face-to-face collaborative sessions. To this end, we performed a user evaluation to compare four different conditions, in which we varied reflections of both workspace and remote people representation. Results suggest potentially more benefits to remote collaboration from workspace consistency rather than people's representation fidelity. We contribute a novel design space, the Negative Space, for remote face-to-face collaboration focusing on 3D content.

## Author Keywords
Telepresence; Workspace Awareness; Face-to-face Communication; Collaborative Systems

## INTRODUCTION
Videoconferencing systems using real size portrayal of people become closer to a co-located experience. In fact, studies suggest that full-body face-to-face communication improves task completion time, presence, and efficiency of communication [12, 7], while enabling non-verbal visual cues including posture, proxemics and deictic gestures in addition to the normal speech and facial expressions currently supported by commercial approaches. Hence, people should rely on natural communication, verbal and non verbal, to convey the focus of the collaboration and pinpoint details on shared content as if they were physically co-located.

When designing for face-to-face collaboration it is necessary to take into account how to address interactions in a shared task space. Despite being typically considered separated from the person space, Ishii et al. [2] suggest that both task and person spaces should be integrated when considering face-to-face meetings using a transparent display metaphor. Indeed, with transparent displays two participants are able to see one another and share digital content, rendered between them, that can be jointly manipulated by both. Yet, in plain face-to-face interactions mediated by displays, people have no common orientation of right or left. Clearboard [2] addresses this issue by mirror-reversing the remote person's video stream, producing gaze and pointing awareness, since 2D graphics and text can thus be corrected to the participant's point-of-view. This approach has been the subject of research for 2D content collaborative manipulation [11, 6, 12]. However, 3D digital content gives rise to detracting issues that affect and impair workspace awareness. Participants do not share the same *forward-backwards* orientation, occlusions can affect the understanding of where or what the remote person is pointing at. Also, contrary points-of-view can result in different perceptions or even serious communication missteps, as illustrated in Figure 1. Naturally, misunderstandings can lead to severe complications. For instance in healthcare, communication errors can cause disastrous consequences when health professionals collaborate to reach a delicate diagnosis or treatment plan, since the analysis of 3D models is commonly used by the medical community [9].

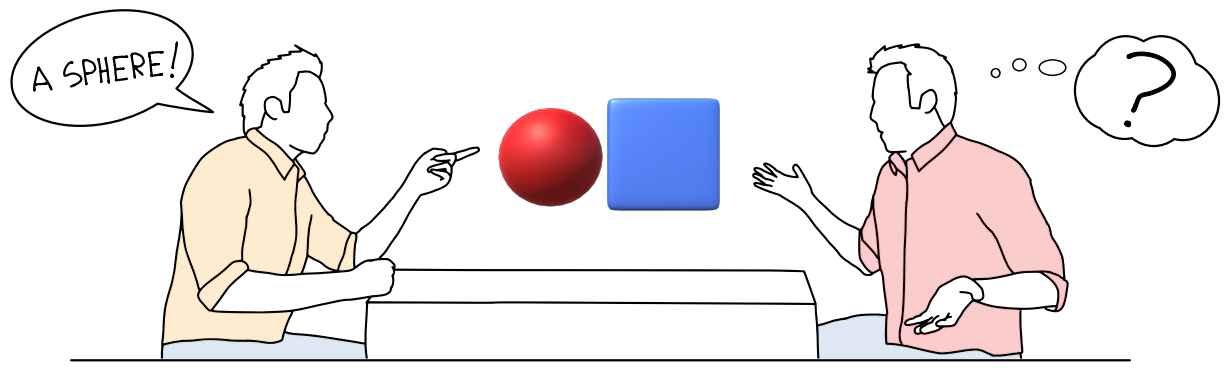

**Figure 1. Example illustration depicting the occlusion issue present when people have opposing points-of-view.**

This work focuses on assessing workspace awareness using variations of the shared workspace settings, individual point-of-view and remote user's representation. For this purpose, we conducted an evaluation comparing task performance and user preferences under four different conditions. We employed an evaluation environment inspired by the *"portal to a distant office"* concept from Wen et al. [10] creating a virtual space between two real spaces. Unlike Wen et al. [10] our collaboration approach provides the three-dimensionality to populate the space between remote rooms with 3D digital content. From the results, we conceptualize the *Negative Space*, an approach to face-to-face remote collaboration, creating a shared virtual workspace linking two physical remote spaces, while providing a sandbox for interacting with 3D content.

## EVALUATING WORKSPACE AWARENESS
We set out to assess if different manipulations of the person-task space can enhance workspace awareness and the way people communicate, when collaborating in a face-to-face setting with 3D content. We developed a full body telepresence prototype and implemented four different workspace conditions. This evaluation was based on both task performance and user preferences. For this, we designed a collaborative 3D assembly task where an *Instructor* guides a remote *Assembler* to reach the correct solution of a toy problem using cubes.

### Evaluation Conditions

To investigate workspace awareness we devised four different evaluation conditions. We performed several combinations of person space and task space. Each combination affects differently the perception of the reference space. Our goal was to study the participants' *point-of-view*, remote participant's *embodiment* and *workspace* rendering, as depicted in Figure 2.

For *point-of-view* we considered that participants could observe workspace in usual opposing points-of-view or simulating an identical viewing experience. Also, similarly to Ishii et al. [3], *embodiment* and *workspace* variables could both be horizontally inverted or not. Handling the permutations of three variables should produce eight different conditions. Though, for this evaluation, we argue beforehand that any combination with an opposing point-of-view other than the real world case, suffers absolutely from worse awareness issues with no improving benefits. Naturally, even in the real world scenario, opposing point-of-views can create communication issues as illustrated in Figure 1, but we decided to maintain it in our evaluation to act as a baseline. We also did not consider the reflected workspace with mirrored embodiment condition, since neither verbal nor deictic gestures match any real reference frame. Therefore, our evaluation followed a within subjects design with four conditions:

1. *Real Life Face-to-face (RL):* Derived from the real world face-to-face scenario, both participants can see each other and the workspace as if they were in opposite ends. As such, the reference space should be natural for the participants since this condition match everyday face-to-face interactions. However, the participants have contrary points-of-view and cannot observe the workspace's opposite side, as demonstrated in Figure 2.A.

2. *Simulated Side-by-side (SS):* While remaining face-to-face in regard to the embodied representation, participants share the same point-of-view of the workspace, in a way that simulates a side-by-side approach. Participants can perceive the workspace from the same side and use verbal relative directions, but pointing gestures from the instructor do not match the reference space (Figure 2.B).

3. *Mirrored Person (MP):* Participants share the same point-of-view, yet the instructor's embodied representation is horizontally inverted to match the reference space. Despite the assembler perceives a mirror embodiment of the instructor, both deictic gestures and verbal relative directions match, as depicted in Figure 2.C.

4. *Mirrored Workspace (MW):* With an identical point-of-view, participants also share faithful face-to-face embodiment representations of each other, although the assembler's workspace is horizontally inverted. Thus, deictic gestures can be used to reference a point. Yet, participants have to accommodate the fact that any verbal relative direction is in reverse (Figure 2.D).

### Method

Participants were grouped in pairs and were asked to perform a set of four tasks, one with each condition, where one participant played the role of the Instructor and the other the Assembler. After completing the four tasks, they were asked to switch roles for another four tasks. All sessions followed the same structure and lasted approximately 50 minutes in total: 10 for the introductory briefing and 20 minutes for each set of four tasks.

We started by introducing the experiment procedure to each pair of participants, followed by a brief description of the interaction technique. Each participant was then randomly assigned his initial role, and was conducted to each individual room. Afterwards, participants jointly executed the evaluation tasks, described in the next session, each task with a different workspace condition.

All tasks were preceded by a training session to familiarize participants with the current condition, and where the assembler had the opportunity to learn how to select and move the checkerboard's blocks. Also, to avoid biased results, in all evaluation sessions the order of the workspace conditions and tasks were performed in an alternated order. Moreover, we devised eight different puzzles to assure that no pair of participants would experience the same puzzle twice. At the end of each task, both participants were asked to fill up a user preferences questionnaire. The evaluation session concluded with profiling questionnaires.

### Tasks

All tasks consisted in solving a block-based puzzle with five colored cubes on top of a checkerboard, where the instructor helps the assembler completing the puzzle using verbal and non-verbal communication cues. That is, instructor's actions and gestures were not augmented by technology and he was not allowed to interact with the virtual task space. For this, a step-by-step description on how to reach the puzzle's solution was provided. Figure 3 shows the multiple steps to complete a puzzle and the position of the instructions' screen. Also, only the instructor could see the colors of the cubes, while for the assembler all were rendered in a neutral gray color, as shown in Figure 4.

All tasks started with the first cube already placed in the correct initial position, while the remaining were randomly placed on the corners of the checkerboard. The instructor's duty was to make it clear to the assembler which cube to pick up next and where to place it, using speech and gestures. The interaction between both participants concluded when all cubes were correctly positioned according to the puzzle's solution. The screen was purposely faded to black signaling the successful ending of a task. To ensure the same complexity between all tasks, the designed puzzles' solutions followed the same set of rules. This was also applied to the training task.

### Setup and Prototype

The evaluation environment consisted of two identical setups replicated in physically separated rooms. Each setup was comprised of an interactive surface (a 55 inch display in portrait mode), two Microsoft Kinect v2, one mounted on top of the surface facing the participant and another one for calibration purposes, and the instructor had a Samsung S7 smartphone for input, as shown in Figure 4.

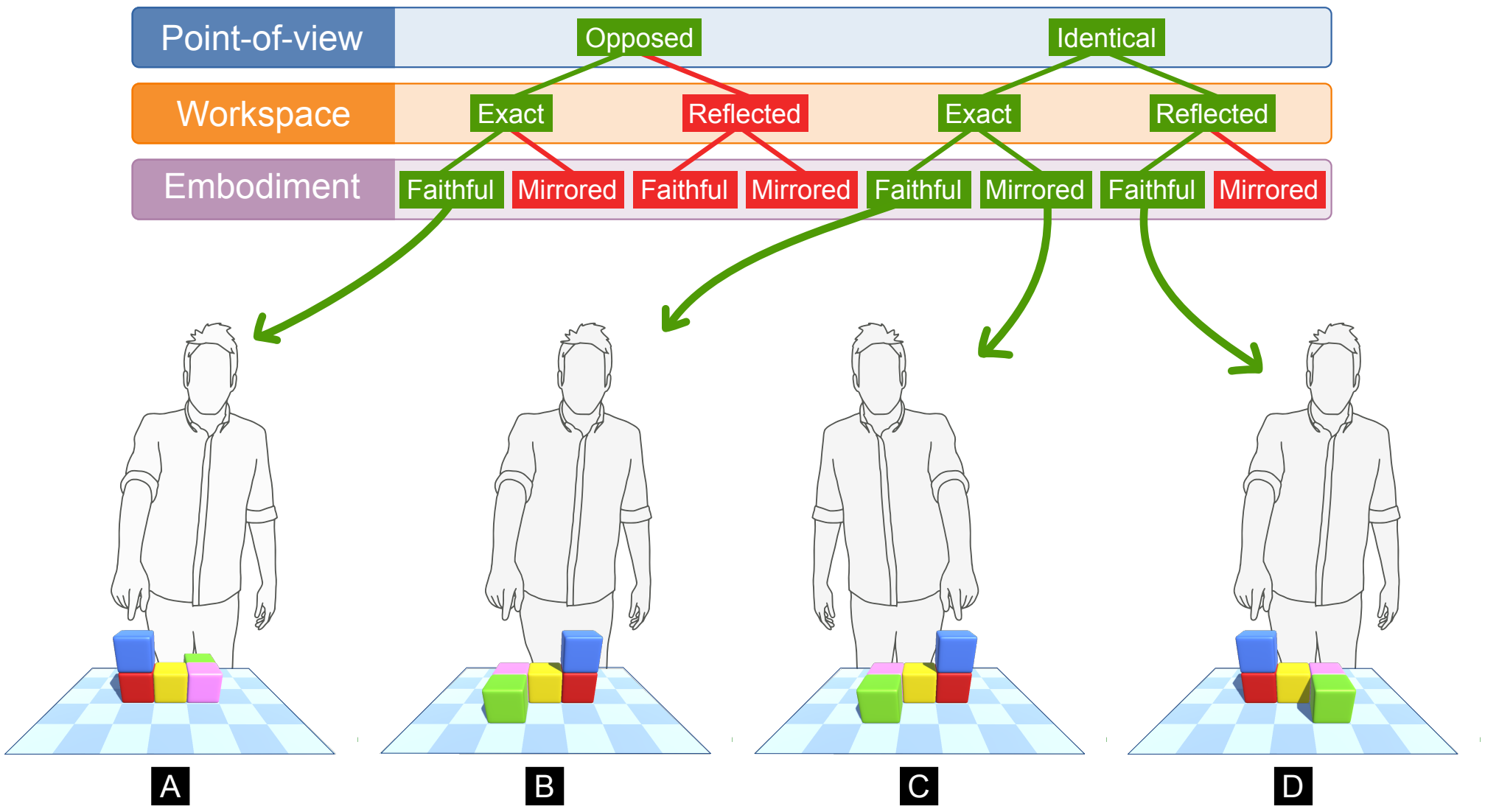

**Figure 2. Workspace conditions from the Assembler's perspective: A) Real life face-to-face (*RL*), B) Simulated side-by-side (*SS*), C) Mirrored Person (*MP*), and D) Mirrored workspace (*MW*). In all cases, the Instructor is pointing with his right index finger to the blue cube.**

We employed a non-intrusive open-source user tracker [8] to combine the interactive surface and the user's body positional data into the same coordinate system. The user's virtual representation was composed of a 3D mesh using color and depth values obtained from the front facing sensor. The prototype was developed in Unity3D and both separated setups were connected in using a network server using TCP connections for the remote user's representation and workspace synchronization. Also, UDP was used to stream the smartphones orientation and button click events over Wi-Fi.

The virtual workspace was constructed inside the closed geometric volume connecting both interactive surfaces to simulate a tunnel between them. We employed a depth of 50 centimeters to accommodate the evaluation's checkerboard. To preserve the illusion of a tunnel between both spaces, the local and remote coordinate spaces were combined, so distances between instructor and assembler were spatially correlated to real distances. We also employed motion parallax by combining a perspective projection [5] with the user's head position retrieved from the user tracker. Using a single shared coordinate system to preserve the real distances and motion parallax promotes the necessary depth cues to convey meaning for the participants' non-verbal cues and deictic gestures. We used the Kinect's own array of microphones and the display's built-in audio speakers to establish audio communication.

The checkerboard was placed on top of the lower plane of the workspace, as shown in Figure 4. To move the board's blocks, we resorted to a *pick* and *drop* approach using a cursor on the screen plane. The assembler could only select one cube at a time by pointing at it. Highlights were activated when the assembler was selecting an object. Selection highlights were shared between both participants. For pointing, we employed the Laser technique [4] using the spatial position of the participant's hand given, combined with the smartphone's orientation from the built-in gyroscope sensor. Then using a ray cast approach, the system was able to determine where the participant was pointing at. Also, the cursor on the screen utilized the participant head to determine the position on the screen plane, to appear on top of the intended location in the workspace.

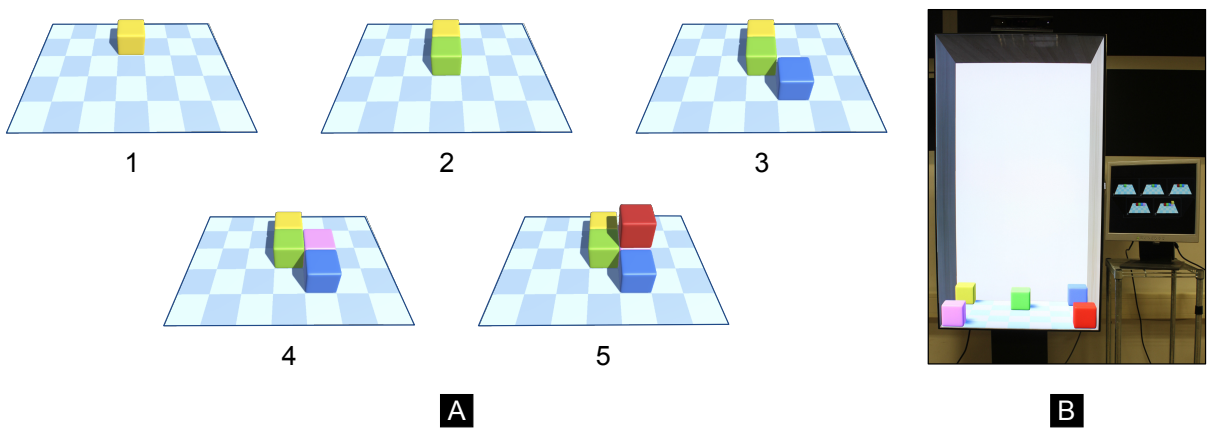

**Figure 3. *Instructors* had access to A) step-by-step solutions for each task in B) a separate display.**

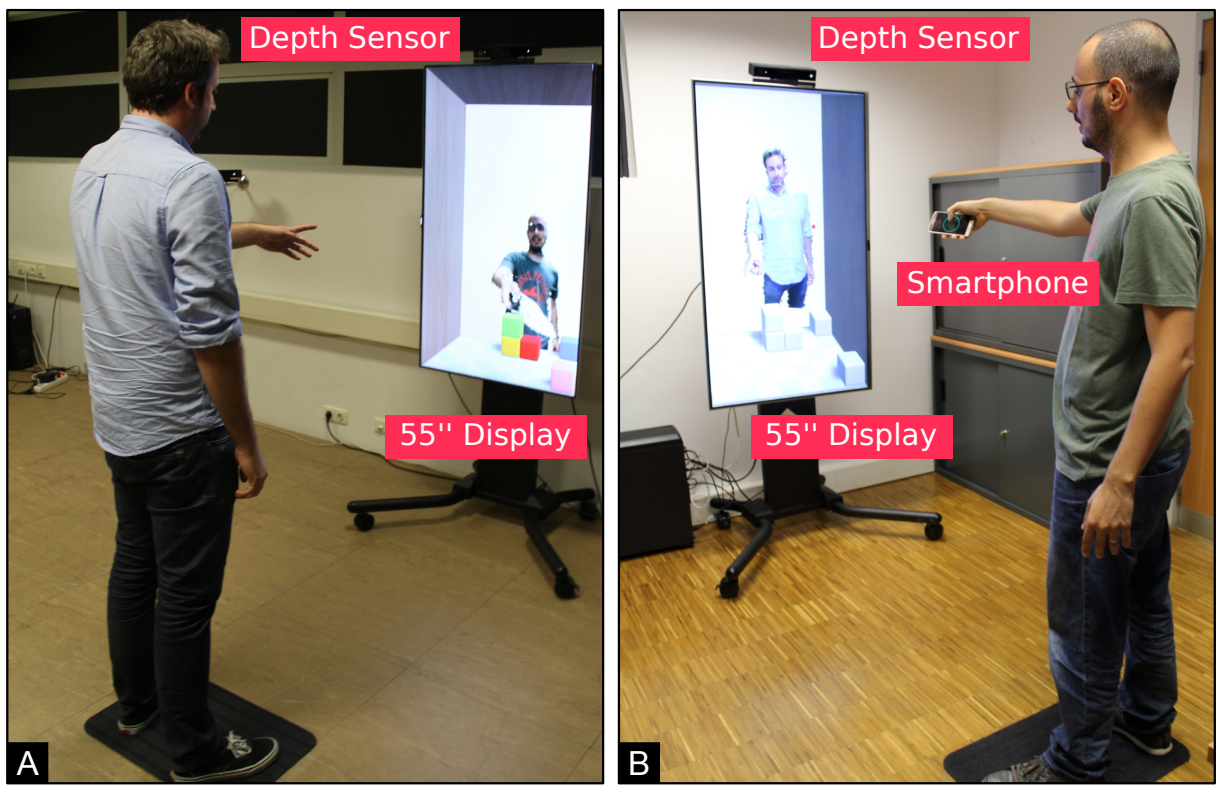

**Figure 4. For the evaluation trials, A) *Instructor* and B) *Assembler* shared similar setups replicated in two separate rooms.**

### Apparatus and Participants

The evaluation trials were performed in two separate rooms in a controlled laboratory environment. Each participant was accompanied by an evaluation moderator. They were instructed to start all tasks on top of a floor mat (Figure 4), positioned at one meter from the display, to preserve the fidelity quality of the participant's embodiment. Despite that, they could freely move around when executing the evaluation tasks.

Our evaluation counted with 16 participants, divided into eight pairs of people. From which, 12 participants were male and four female, and with the great majority (approximately 94%) were between 18 and 35 years old. Most had at least a BSc degree (89%). All participants did not exhibit any color vision deficiency after performing a standard Ishihara test [1] with nine different plates.

## RESULTS

During the evaluation trials we collected *Task Performance* data through logging, and gathered *User Preferences* from questionnaires filled up after the execution of each task. To perform the statistical analysis, we firstly used Shapiro-Wilk test to assess data normality. For the evaluation conditions, we ran the repeated measures ANOVA test to find significant differences in normal distributed data, and Friedman non-parametric test with Wilcoxon-Signed Ranks post-hoc test otherwise. To test for influence of puzzle complexity in tasks' performance, we resorted to One-Way ANOVA. In all cases, post-hoc tests used a Bonferroni correction.

### Task Performance Overview

We logged completion times, number of wrong cube selections and wrong cube placements. To certify that all eight puzzles presented similar complexity, we tested their times with a One-Way ANOVA, which showed that they were indeed similar (F(7,53)=1.426, p=.215), meaning that different puzzles did not affect in any way the task performance. Figure 5 shows tasks' completion times for each condition. Although it appears to be a tendency for lower times with the *MP* condition, no statistically significant differences between the four workspace conditions were found (F(2.218,28.831)=1.981, p=.152). Wrong cube selections and placements are reported in Table 1. Again, no statistically significant differences were found for either wrong selections ($\chi^2$(3)=1.719, p=.633) or placements ($\chi^2$(3)=2.038, p=.565).

| | Wrong Selections | Wrong Placements |
|---|---|---|
| *RL* | 0.5 (1.25) | 2 (1.5) |
| *SS* | 0 (0.75) | 1 (2) |
| *MP* | 0 (0.25) | 1 (1.25) |
| *MW* | 0 (1.25) | 1 (2) |

* indicates statistical significance

**Table 1. Tasks' number of wrong selections and placements (Median, Inter-quartile Range).**

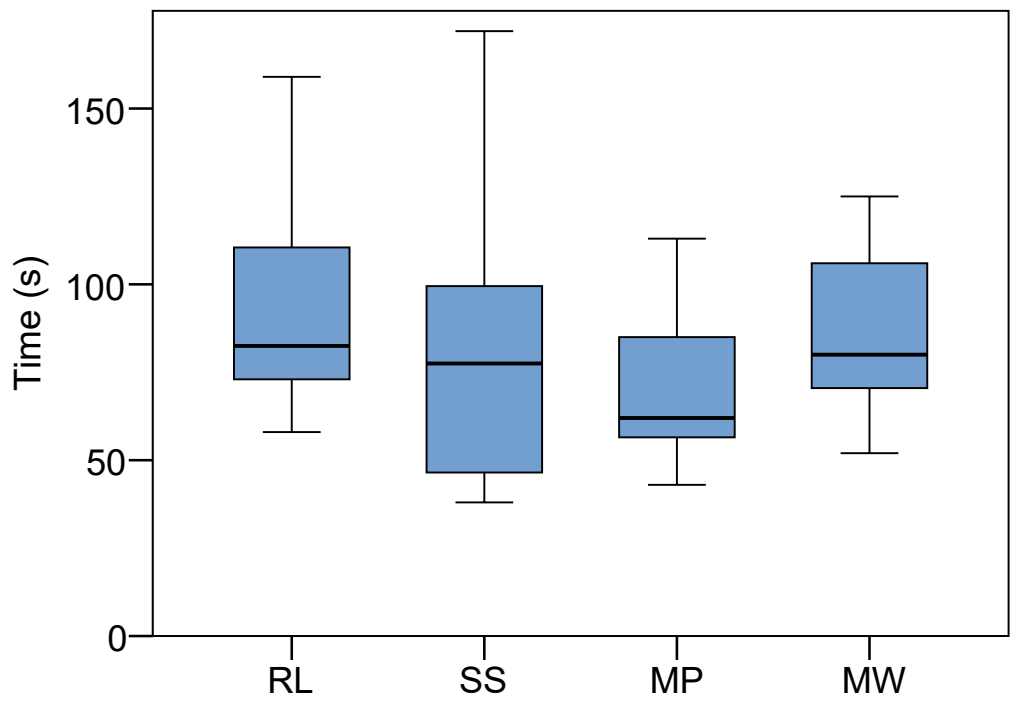

**Figure 5. Tasks' completion times for each condition.**

### User Preferences Overview

After the completion of each task, participants were asked to fill up a preferences questionnaire related to the condition they just experimented. Table 2 shows prompted questions on the user preferences questionnaire with results for both assembler and instructor in all four conditions. Statistical significant differences were found on three questions for the instructor (Q1:($\chi^2$(3)=10.892, p=.012); Q3: ($\chi^2$(3)=11.598, p=.009); Q4: ($\chi^2$(3)=10.102, p=.018)). The post-hoc test revealed that participants in the instructor's role strongly agreed that *MW* was more difficult than *SS* (Z=-2.743, p=.006) and *MP* (Z=-2.722, p=.006). Instructors agreed that in the *MP* condition, explaining the row of the cube to select, is easier than *MW* (Z=-2.967, p=.003). It was also easier for instructors to explain the column of the next cube to be selected in the *MP* condition than *MW* (Z=-2.675, p=.007). We did not find any significant statistical difference after participants experienced tasks in the role of assembler.

### Observations

During the execution of each task, we observed the participants' behavior regarding their communication style and usage of gestures. Although personal and cultural differences can influence this result, we were able to identify certain trends in each condition. Throughout all conditions, verbal communication was predominant using combined spatial and temporal references (e.g. *"left to the cube you have previously moved."*). We observed that participants developed an informal shared protocol to better understand how to complete the task. This was achieved by the instructor asking several questions to the assembler. More specifically, instructors inquired if the assembler could raise a arm and/or select a cube on a specific corner of the workspace. Henceforth, instructors would communicate the commands already in the assemblers' reference frame, which justifies the existence of significant differences in the questionnaires only for instructors.

Participants that started with *RL* condition used indicative gestures much more naturally and frequently, until experiencing the *SS* where these were ambiguous. At that point, the mentioned communication style would be established, overpowering deictics, which would be only applied as a last resort. Even so, involuntary non-verbal cues such as gaze, subtle hand, finger gestures accompanying speech, or leaning the body to

| It was easy to... | Instructor RL | Instructor SS | Instructor MP | Instructor MW | Assembler RL | Assembler SS | Assembler MP | Assembler MW |
|---|---|---|---|---|---|---|---|---|
| [Q1]...complete the task.* | 5.5 (1) | 6 (1) | 6 (1) | 5 (1) | 5 (1.25) | 6 (1) | 5 (0.25) | 5 (2) |
| [Q2]...explain/understand which cube to select. | 6 (1) | 6 (1) | 5.5 (1) | 5 (1) | 5 (0.25) | 6 (1) | 5.5 (1) | 5 (2) |
| [Q3]...explain/understand the row of the cube to select.* | 6 (1) | 6 (1.25) | 6 (0.25) | 5 (0.25) | 5 (0.25) | 5.5 (1) | 5 (1) | 5 (2.25) |
| [Q4]...explain/understand the column of the cube to select.* | 5.5 (1.25) | 6 (1) | 6 (0.25) | 5 (2.25) | 5 (0.25) | 6 (1) | 5.5 (1) | 5 (1) |
| [Q5]...explain/understand where to position the cube. | 5 (1.25) | 6 (1) | 5.5 (1) | 5 (1.25) | 5 (1.25) | 5.5 (1) | 5 (1.25) | 5 (1) |
| [Q6]...explain/understand the row to position the cube. | 5 (1) | 6 (1.25) | 6 (1.25) | 5 (1.25) | 5 (2) | 5.5 (1) | 5 (0.5) | 5 (1.25) |
| [Q7]...explain/understand the column to position the cube. | 5 (2) | 6 (1) | 6 (1) | 4.5 (2.25) | 5 (2) | 5.5 (1) | 5 (1.25) | 4.5 (1.25) |

* indicates statistical significance

**Table 2. Results of the user preferences questionnaires (Median, Inter-quartile Range).**

a certain direction was frequently picked up by assemblers, who would try to predict the next instruction according to these visual cues. Explicit line and column indications had seldom use and had a negative impact in all of its occurrences. Indications such as *"third row, second column"* were harder to disambiguate than temporal references.

The usage of non-verbal communication varied widely according to the workspace condition. In *RL*, gestures were used to disambiguate depth, given that it was the only condition where this mapping was accurate. Also, *RL* was the only condition where we had some users use non-verbal cues as their main communication method. In *SS*, all attempts of using hand gestures resulted in errors by the assemblers. *MP* allowed users to use gestures naturally as a complement to clear verbal instructions. Finally, in *MW*, gestures were used by majority of participants, but less accurately than in *RL*, due to the fact that there was not a direct mapping between pointing and verbal directions.

### Discussion

Results show an absence of significant differences in task performance. Yet, time data appears to reveal a tendency for *MP* to allow faster task completion. Regarding user preferences, statistical significant differences were found on instructors' answers. This happened because it was mostly the instructor who did the calculations regarding reference frames, which rendered all conditions alike to the assembler.

Although participants established the informal shared protocol to calibrate reference frames and achieved similar performance in all conditions, a *Reflected* workspace was clearly identified as being more difficult than the *Exact* representation. We argue that the cognitive workload of being constantly converting coordinates between both frames is mentally demanding.

In complex scenarios, where it is imperative for both participants to observe the same details, the *RL* condition is unfit. This and the cognitive cost associated to the *MW* condition, leads us to suggest that, for this kind of scenarios, having an *Exact* workspace with an *Identical* point-of-view is highly desirable. The choice between *SS* or *MP* will be dependent on whether the accuracy of the remote person's representation is more relevant than the consistency between the person and task spaces, respectively.

## NEGATIVE SPACE

Previous research on remote face-to-face collaboration have successfully contributed full-body telepresence approaches with integrated person-task spaces, which demonstrated considerable improvements on presence and cooperative task performance. Most focus on cooperative interactions with 2D content, although collaboration in design and review of 3D virtual models is crucial in several domains.

With this in mind, we introduce *Negative Space* as a conceptual platform with a set of rules for future works on remote collaboration. It is characterized as a virtual space that serves as a gateway between two physical rooms where collaborative 3D interactions can occur, as depicted in Figure 6. From the evaluation results presented in the previous section, we devised *Negative Space* as a medium to support discussions on shared views over the 3D content and, as such, it shall offer participants *Identical* points-of-view over *Exact* copies of the workspace. We also enforce the usage of real-time 3D reconstructions of remote people for improved perception. Our approach can be advantageous in avoiding communication breakdowns by making many gestures and deictic idioms easier to share and understand between participants.

Similarly to Ishii et al. [2], *Negative Space* exploits the benefits of a see-through display. By positioning the virtual content between two people, participants are able to profit from normal face-to-face interactions as if they were physically co-located. This contributes to the overall workspace and situational awareness, since participants are able to observe the other person's gaze direction, deictic gestures and actions, while performing selection and manipulation tasks related to multiple occupa-tional fields, such as engineering, industrial, architecture and medical, as demonstrated in Figure 7.

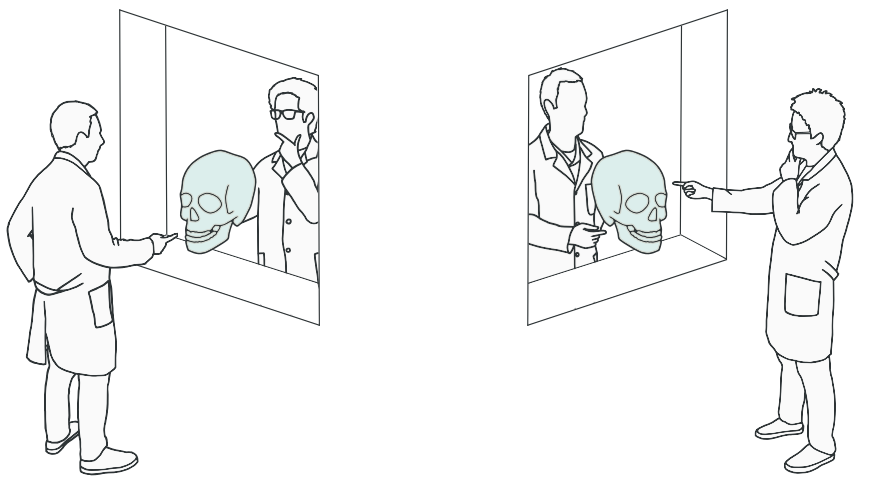

**Figure 6. Conceptual vision of Negative Space featuring two remote locations. Participants have identical points-of-view over exact copies of the workspace.**

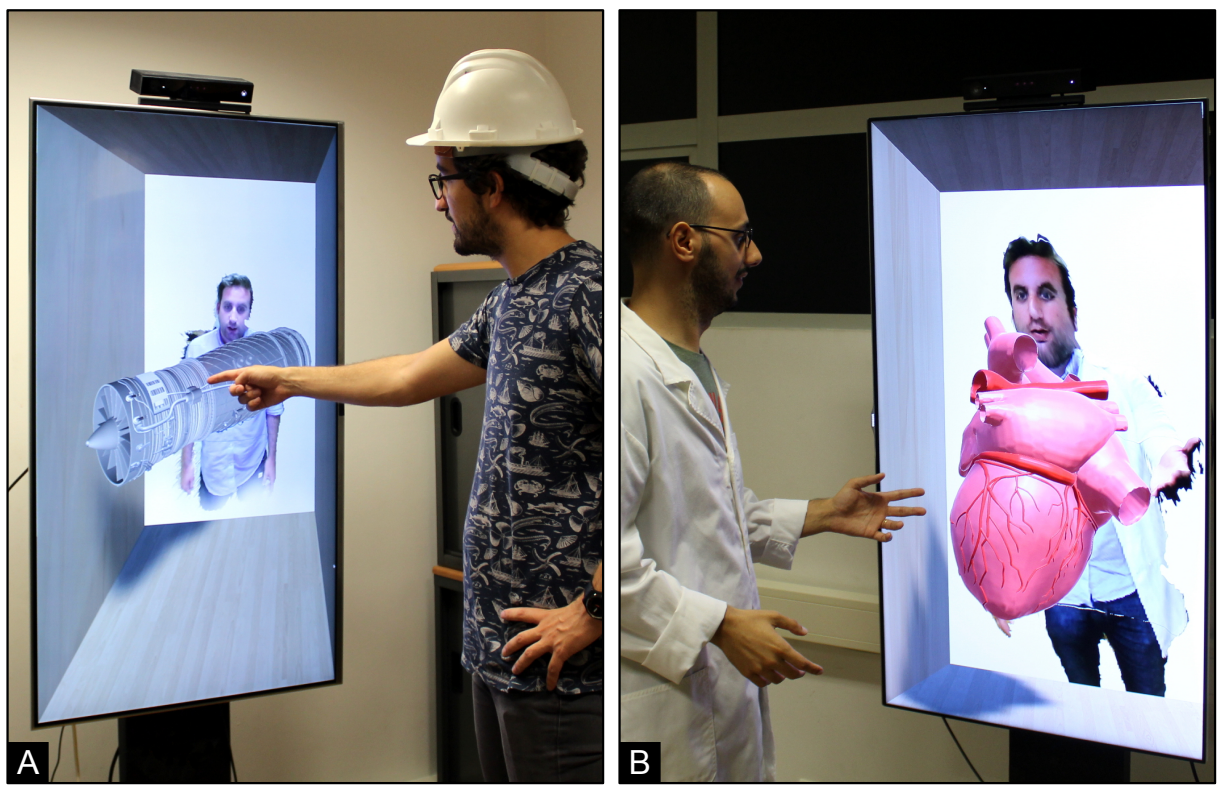

**Figure 7. The *Negative Space* concept can be applied in multiple usage scenarios requiring visualization, design and review of virtual 3D models. Notable examples are A) engineering industries and the B) healthcare.**

## CONCLUSIONS AND FUTURE WORK

In this work we presented an evaluation of several combinations of different points-of-view, and workspace and embodiment characteristics to study remote face-to-face collaborative work on 3D shared content, with the objective of achieving a consistent and seamless reference space between participants while promoting workspace awareness. For this, we devised instructor-assembler trials, where participants jointly solved a puzzle in four conditions. Results show that participants can successfully collaborate in a shared 3D workspace face-to-face, and suggest that having an identical point-of-view is essential. Also, having an exact task space is highly desirable to avoid the cognitive cost of collaborating when remote people cooperate with different views of the shared 3D content.

As a consequence of the results' analysis, we conceptualize *Negative Space*, a telepresence approach that enables full-body face-to-face communication and creates a virtual task space between two remote spaces, where interactions with 3D objects can occur. Our proposed face-to-face approach for collaboration considers that remote participants share an identical view of the same exact task space and are able to perceive one another as if they were in same physical space. We believe that the *Negative Space* can serve as a bootstrap template for future studies and developments on face-to-face collaborative work with 3D objects.

Finally, with our work, we identified avenues for future research. Although mirroring the remote person representation fixes left-right discrepancies, there are still depth mismatch issues that affect the perception of deictic gestures. This could be tackled through body warping and/or pointing manipulation, for instance. Additionally, further studying selection and manipulation techniques for design and review of 3D models can help improve remote collaboration within the *Negative Space*. Lastly, support for groups of more than two geographically distributed people might extend face-to-face collaboration on 3D models in novel and exciting ways.